# The diffraction power of crystals with unknown atomic structures: calculation and application in quantitative phase analysis

Hui Li[a]*, Meng He[bcd]* and Ze Zhang[e]

[a] Institute of Microstructure and Properties of Advanced Materials, Beijing University of Technology, 100 Ping Le Yuan, Chaoyang District, Beijing, 100124, People's Republic of China
[b] CAS Key Laboratory of Nanosystem and Hierarchical Fabrication, CAS Center for Excellence in Nanoscience, National Center for Nanoscience and Technology, Beijing, 100190, People's Republic of China
[c] School of Physical Sciences, University of Chinese Academy of Sciences, Beijing, 100049, People's Republic of China
[d] Tianmu Lake Institute of Advanced Energy Storage Technologies, Liyang, Jiangsu Province, 213300, People's Republic of China
[e] Zhejiang University, Hangzhou, Zhejiang Province, 310014, People's Republic of China

Correspondence email: huilicn@yahoo.com; mhe@nanoctr.cn

**Funding information** National Natural Science Foundation of China (Grant No. 11574060).

**Synopsis** A method was developed to calculate the diffraction power of crystals with an unknown atomic arrangement. Consequently, quantitative phase analysis using X-ray powder diffraction data can be made for samples including phases with unknown atomic structures.

**Abstract** Quantitative phase analysis is one of the major applications of X-ray powder diffraction. The essential principle of quantitative phase analysis is that the diffraction intensity of a component phase in a mixture is proportional to its content. Nevertheless, the diffraction intensity of the component phases cannot be compared with each other directly since each phase has its specific diffraction power. The diffraction power of the unit cell of a crystal is well represented by the sum of the squared structure factors, which cannot be calculated directly when the structure data is unavailable. Here a method was demonstrated to calculate the diffraction power using only the chemical contents in the unit cell of a crystal, which enables quantitative phase analysis on a mixture sample consisting of crystalline phases with unknown atomic structures.

1. **Introduction**

As a powerful tool for phase identification and quantitative phase analysis, X-ray powder diffraction has been extensively applied in both research and industry. The essential principle for quantitative phase analysis is that the diffraction intensity of a component phase is proportional to its content in the mixture. Nevertheless, different crystalline phases have various diffraction power. Therefore, the diffraction intensity of the component phases in a mixture cannot be compared with each other directly, and then related to the content of a component phase. Indeed, the diffraction intensity of each phase in the mixture has to be normalized by its diffraction power before it can be related to the content of the corresponding phase.

Structure factor is the ratio of the amplitude of the diffracted wave of the unit cell to that of the free electron at the origin of the unit cell. The intensity of a certain reflection is proportional to the squared structure factor when experimental parameters, such as Lorenz-polarization factors, were not taken into accounts. Then, the sum of the squared structure factors over all possible reflections can well represent the diffraction power of a crystal. The structure factor can be readily calculated when the atomic arrangement in the unit cell is known:

$$F_h = \sum f_i(h) \exp(i2\pi \boldsymbol{h} \cdot \boldsymbol{r_i}) \qquad (1)$$

where $F_h$ is the structure factor, $f_i$ is the atomic scattering factor of the $i$th atom in the unit cell, $\boldsymbol{h}$ is the diffraction vector, and $\boldsymbol{r}$ is the positional vector of the $i$th atom. The diffraction power of the unit cell of a certain crystal, $\sum_h |F_h|^2$, is then determined. If the atomic structure data is available for all the phases in the mixture, the content of each component phase can be

deduced from the observed diffraction intensity normalized by the diffraction power of the corresponding phase.

The atomic structure data is not always available for each phase in the mixture to be quantitatively analyzed. Many methods have been developed and extensively applied for such samples, such as the absorption-diffraction method, the spiking method, and the internal standard method, as summarized by Pecharsky and Zavalij (Pecharsky and Zavalij, 2009). With these methods, one does not need the atomic structure of the component phases to deduce the content of each phase from the observed X-ray powder diffraction data. Nevertheless, additional samples, such as the pure phase of the target component and the investigated mixtures spiked with a known amount of the target phase or standard reference materials, have to be prepared, and each of them has to be measured with X-ray powder diffraction. Therefore, the process of quantitative phase analysis based on the above methods usually is experimentally tedious and time-consuming. A concept of reference intensity ratio (RIR) was developed to simplify the process of quantitative phase analysis (Chung, 1974). When a crystalline phase was mixed with common reference material (usually corundum) with a weight ratio of 1:1, then the intensity ratio of the strongest reflection of each phase was defined as RIR. If the RIR of each component phase in a mixture is known, then the concentration of a component phase can be deduced directly from the intensity ratio between the target phase and the spiked reference material. Unfortunately, RIR is not always available for phases to be quantitatively analyzed. It is especially true for new phases encountered in research and development activities. More importantly, the intensity ratio between the target phase and the reference material depends on the specimen preparation and experimental parameters of the X-ray diffraction data collection. Therefore, the RIR listed in the ICDD PDFs may be different greatly from the true values exhibited in one's experimental work. In addition, the RIR method utilizes only the intensity information of the strongest reflection, which is very sensitive to the specimen preparation and experimental parameters of data collection. Toraya (Toraya, 2016) proposed a method for quantitative phase analysis in 2016, which derives the content of the component phase from the diffraction intensities and the chemical composition of the target phase. This method enables the quantitative phase analysis for samples including phases with unknown atomic structures while no additional auxiliary samples and diffraction datasets are necessary. In this method, the diffraction power of the unit cell of a component phase was estimated using the product of the unit cell volume and the sum of the squared electron numbers of each atom over the whole unit cell. Namely, the diffraction power was assessed by

$$\sum_h |F_h|^2 = CV \sum_{i=1}^{N} n_i^2 \qquad (2)$$

where $C$ is a proportional constant, $V$ is the unit cell volume, $n_i$ is the electron number of the $i$th atom in the unit cell, $N$ the total number of atoms in the unit cell. Unfortunately, the equation (2) was derived from an assumption that the peak height of the Patterson function at the origin can be approximated by the integrated convoluted electron density of the peak. This assumption has no solid theoretical foundation and logical proof. Moreover, to apply equation (2) in quantitative phase analysis, the proportional constant $C$ has to be assumed to be common to and independent of the various component phases in the investigated mixture. Actually, there is no scientific evidence for the assumption of "constant C".

Here we present a new method to calculate the unit cell diffraction power of crystalline phases with unknown atomic structures. All information needed to implement the calculation is the unit cell and its chemical contents. The approach to quantitatively analyze the content of a structurally unknown component phase in a mixture has been developed based on our new method of the diffraction power calculation. The validity of both the method of diffraction power calculation and the approach of quantitative phase analysis was demonstrated.

## 2. Theory

**2.1. Calculation of diffraction power using the chemical contents of the unit cell**

The structure factor, $F_h$, of a crystal with an electron density distribution of $\rho(r)$ in its unit cell is the Fourier transform of $\rho(r)$, namely,

$$F_h = \int_V \rho(r) \exp(i2\pi h \cdot r) \, dv \qquad (3)$$

where $V$ is the volume of the unit cell, $r$ is the positional vector, $h$ the diffraction vector.
Applying Parseval theorem to equation (3), then we have

$$\sum_h |F_h|^2 = \int_V \rho^2(r) dv \qquad (4)$$

In a crystal, the electron density distribution of a constituent atom will be slightly different from that of a free atom of the same species due to the formation of chemical bonding. Nevertheless, the difference is so small that it can hardly be detected by a regular X-ray powder diffraction measurement. Thus, for X-ray diffraction measurement, the electron density distribution in the unit cell of a crystal may be approximately taken as the sum of the electron density of a series of free atoms, each of which is of the same species and position as the corresponding constituent atoms in the unit cell of the crystal. Namely,

$$\rho(\mathbf{r}) = \sum_{i=1}^{n} \rho_i(\mathbf{r_i}) \tag{5}$$

where n is the total number of atoms in the unit cell, $\rho_i(\mathbf{r_i})$ is the electron density distribution of a free atom, which is of the same species and position as the $i$th constituent atom in the unit cell. Since the overlap of electron density between adjacent atoms is negligible, we have

$$\int_V \rho^2(\mathbf{r}) dv = \int_V [\sum_{i=1}^{n} \rho_i(\mathbf{r_i})]^2 dv = \int_V \sum_{i=1}^{n} \rho_i^2(\mathbf{r_i}) dv = \sum_{i=1}^{n} \int_V \rho_i^2(\mathbf{r_i}) dv$$

(6)

Let us consider an imaginary crystal, which has the same unit cell as the crystal under the investigation. There is only one atom in the unit cell of the imaginary crystal, and the only atom is of the same species as the $i$th atom in the unit cell of the crystal under the investigation and located at the origin of the unit cell of the imaginary crystal. Then the structure factor of the imaginary crystal is given by

$$F_{i,\mathbf{h'}} = \sum f_i(\mathbf{h'}) \exp(i2\pi \mathbf{h'} \cdot \mathbf{r'}) = f_i(\mathbf{h'}) \equiv f_{i,h'} \tag{7}$$

or

$$F_{i,\mathbf{h'}} = \int_V \rho_i(\mathbf{r'}) \exp(i2\pi \mathbf{h'} \cdot \mathbf{r'}) dv \tag{8}$$

where $F_{i,\mathbf{h'}}$ is the structure factor of the imaginary crystal, $\mathbf{h'}$ and $\mathbf{r'}$ are the diffraction vector and positional vector of the imaginary crystal, respectively, $f_i(h')$ is the atomic scattering factor of the constituent atom of the imaginary crystal, and $\rho_i(r')$ the electron density distribution in the unit cell of the imaginary crystal.

Applying Parseval theorem to equation (8), then we have

$$\sum_{h'} |F_{i,h'}|^2 = \int_V \rho_i^2(\mathbf{r'}) dv \tag{9}$$

Combining equation (5) and (7), we have

$$\int_V \rho_i^2(\mathbf{r'}) dv = \sum_{h'} f_{i,h'}^2 \tag{10}$$

The only difference between $\rho_i(\mathbf{r'})$ and $\rho_i(\mathbf{r_i})$ is a positional translation. Then combining equation (4), (6) and (10), we have

$$\sum_h |F_h|^2 = \sum_{i=1}^{n} \sum_{h'} f_{i,h'}^2 \tag{11}$$

Using equation (11), we can calculate the diffraction power of the unit cell of a crystal without knowing the atomic arrangement in the unit cell. All we need to perform the calculation is the unit cell and its chemical contents.

2.2. **Application in the quantitative phase analysis**

In X-ray powder diffraction, the intensity of reflection $h$ of the $j$th component phase in a mixture consisting of J phase is given by

$$I_{j,h} = K \frac{v_j}{V_j^2} G_{j,h} |F_{j,h}|^2 \qquad (12)$$

where $I_{j,h}$ is the intensity of reflection $h$ of the $j$th component phase, K is a proportional factor, $v_j$ is the volume fraction of the $j$th phase, $V_j$ is the volume of the unit cell of the $j$th phase, $G_{j,h}$ is the parameter related to the diffraction geometry, and also dependent on $h$. $G_{j,h}$ can be calculated when the diffraction geometry and the lattice parameters of the $j$th phase are known. $F_{j,h}$ is the structure factor of the $j$th phase.

Then the sum of the diffraction intensity of the $j$th phase is given by

$$\sum_h I_{j,h} = \sum_h K \frac{v_j}{V_j^2} G_{j,h} |F_{j,h}|^2 \qquad (13)$$

So that the volume fraction of the $j$th phase can be derived:

$$v_j = \frac{V_j^2 \sum_h I_{j,h}/G_{j,h}}{K \sum_h |F_{j,h}|^2} \qquad (14)$$

According to equation (11), the volume fraction of the $j$th phase can also be calculated using the chemical contents in the unit cell instead of the structure factors:

$$v_j = \frac{V_j^2 \sum_h I_{j,h}/G_{j,h}}{K \sum_{i=1}^{n} \sum_{h'} f_{j,i,h'}^2} \qquad (15)$$

where $f_{j,i,h'}$ is the atomic scattering factor of the $i$th atom of the $j$th phase.

In equation (14) and (15), K is a parameter to be determined. It can be derived from

$$\sum_{j=1}^{J} v_j = 1 \qquad (16)$$

When K is determined, the volume fraction, $v_j$, of each phase in the mixture can be derived using equation (14) or (15). The volume fraction can be readily converted to weight fraction by

$$w_j = \frac{\dfrac{M_j Z_j V_j \Sigma_h (\frac{I_{j,h}}{G_{j,h}})}{\Sigma_{i=1}^{n} \Sigma_h f_{j,i,h}^2}}{\Sigma_{j'}^{J} \dfrac{M_{j'} Z_{j'} V_{j'} \Sigma_{h'}(\frac{I_{j',h'}}{G_{j',h'}})}{\Sigma_{i'=1}^{n'} \Sigma_{h'} f_{j',i',h'}^2}} \qquad (17)$$

where $w_j$ is the weight fraction of the *j*th phase, $M_j$ and $Z_j$ are the chemical formula weight and the number of chemical formula in the unit cell of the *j*th component phase, respectively.

The weight fraction of each component in the mixture can be calculated using equation (17), while all information needed to perform the calculation is the unit cell and the chemical contents in the unit cell of each phase in addition to an X-ray powder diffraction pattern of the mixture. Neither pure phase of the component phase, reference materials, additional auxiliary samples and diffraction datasets nor atomic structure data, or RIR information is necessary. Of course, if atomic structure data is available for some components, $\Sigma_h |F_h|^2$ in place of $\Sigma_{i=1}^{n} \Sigma_{h'} f_{i,h'}^2$ can be calculated and used in equations (15) and (17) for these phases.

3. **Validation and discussion**

3.1. The consistency between $\Sigma_h |F_h|^2$ and $\Sigma_{i=1}^{n} \Sigma_{h'} f_{i,h'}^2$

Theoretically, equation (11) is valid only when the overlapped electron density of adjacent atoms is negligible and all possible reflections are taken into account. Actually, it is a reasonable approximation for a regular X-ray diffraction measurement that the overlapped electron density is negligible. Nevertheless, only reflections below a certain upper limit of the Bragg angle can be measured in a practical X-ray diffraction measurement. To apply equation (11) in analyzing the practical X-ray diffraction data, its validity has to be checked when only reflections in a limited Bragg angle range are available. Here we calculated both $\Sigma_h |F_h|^2$ and $\Sigma_{i=1}^{n} \Sigma_{h'} f_{i,h'}^2$ for several crystalline phases, namely Si, NaCl, α-$Al_2O_3$, $Li_2CO_3$ and $Ag_2Te$ (Hessite). The atomic structure data of these phases are obtained from the literature. The wavelength corresponding to Cu Kα radiation was assumed and the upper limit of Bragg angle (2θ) was set to be 60, 80, 100, 120 and 140 °. The results are presented in Table 1. When atomic displacement parameters are not taken into account, namely B = 0 is assumed, a discrepancy in the range of 10% to 15% was observed between $\sum_h |F_h|^2$ and $\sum_{i=1}^n \sum_{h'} f_{i,h'}^2$ at the upper limit of Bragg angle of 60 ° for all phases except for NaCl, for which the discrepancy is as low as about 1%. When the upper limit of 2θ is 80 ° or higher, generally, much smaller inconsistencies between $\sum_h |F_h|^2$ and $\sum_{i=1}^n \sum_{h'} f_{i,h'}^2$ are observed. The largest discrepancy, 11.6%, was observed for α-Al$_2$O$_3$ at the upper limit of 2θ of 120 °, while in most cases the discrepancy is below 10%. As evidenced by the case study, $\sum_{i=1}^n \sum_{h'} f_{i,h'}^2$ is a reasonable approximation of $\sum_h |F_h|^2$ when a sufficient number of reflections are taken into account. For a regular X-ray powder diffraction measurement using Cu Kα radiation, an upper limit of 2θ of 80 ° seems to be adequate for validating equation (11). We also noted that the ratio of $\sum_{j=1}^n \sum_{h'} f_{j,h'}^2 / \sum_h |F_h|^2$ fluctuate with the increase of the upper limit of 2θ until 140 °, rather than converge to 1. This results from the fact that reflections of crystals are distributed in the observed 2θ range discretely and irregularly. It is also noteworthy that the ratios of $\sum_{j=1}^n \sum_{h'} f_{j,h'}^2 / \sum_h |F_h|^2$ observed for Li$_2$CO$_3$ and α-Al$_2$O$_3$ are always greater than 1, while the ratios observed for other phases fluctuate around 1. Compared with other phases, Li$_2$CO$_3$ and α-Al$_2$O$_3$ consist of atoms with a lower average atomic number. It is not clear if this is just an occasional phenomenon or if there is some regularity behind it.

**Table 1** $\sum_{j=1}^n \sum_{h'} f_{j,h'}^2 / \sum_h |F_h|^2$ calculated for several crystalline phases with the assumption of atomic displace parameters B = 0 and 1.5 Å$^2$, respectively.

| phases | Space group | $\sum_{j=1}^n \sum_{h'} f_{j,h'}^2 / \sum_h |F_h|^2$ | | | | | |
|---|---|---|---|---|---|---|---|
| | | B (Å$^2$) | Upper limit of 2θ (°) | | | | |
| | | | 60 | 80 | 100 | 120 | 140 |
| Si | $Fd\bar{3}m$ | 0 | 0.872 | 1.073 | 1.025 | 1.000 | 0.990 |
| | | 1.5 | 0.905 | 1.080 | 1.041 | 1.023 | 1.016 |
| NaCl | $Fm\bar{3}m$ | 0 | 1.011 | 1.057 | 1.112 | 0.958 | 1.019 |

|  |  | 1.5 | 1.027 | 1.071 | 1.097 | 0.998 | 1.034 |
|---|---|---|---|---|---|---|---|
| α-Al$_2$O$_3$ | $R\bar{3}c$ | 0 | 1.117 | 1.026 | 1.102 | 1.116 | 1.043 |
|  |  | 1.5 | 1.239 | 1.083 | 1.135 | 1.142 | 1.096 |
| Li$_2$CO$_3$ | $C2/c$ | 0 | 1.100 | 1.090 | 1.066 | 1.018 | 1.022 |
|  |  | 1.5 | 1.123 | 1.106 | 1.092 | 1.059 | 1.060 |
| Ag$_2$Te (hessite) | $P2_1/c$ | 0 | 1.149 | 0.974 | 1.079 | 0.999 | 0.989 |
|  |  | 1.5 | 1.146 | 0.997 | 1.071 | 1.019 | 1.011 |

**3.2. The effect of atomic displacement parameters on the consistency**

It is well known that atoms in a real crystal are actually vibrating about their equilibrium positions, and the atomic displacements reduce the atomic scattering factors. Then both sides of equation (11) are affected. We took the atomic displacement into account, and re-calculated $\sum_h |F_h|^2$ and $\sum_{i=1}^n \sum_{h'} f_{i,h'}^2$ reported in section 3.1 to illustrate the effect of atomic displacement on the consistency between them. A typical atomic displacement parameter, B = 1.5 Å$^2$, was assumed for all atoms. The recalculated values were also presented in Table 1 in comparison with the results obtained with the assumption B = 0 Å$^2$. As shown in Table 1, generally, the consistency between $\sum_h |F_h|^2$ and $\sum_{i=1}^n \sum_{h'} f_{i,h'}^2$ does not improve when B = 1.5 Å$^2$ was assumed for all atoms. Actually, in most cases, the ratios of $\sum_{j=1}^n \sum_{h'} f_{j,h'}^2 / \sum_h |F_h|^2$ increased in comparison with their counterparts calculated by assuming B = 0 Å$^2$. Based on these limited preliminary results, it seems reasonable to assume B = 0 Å$^2$ when one calculates $\sum_{i=1}^n \sum_{h'} f_{i,h'}^2$ as an approximation of $\sum_h |F_h|^2$.

**3.3. Quantitative phase analysis**

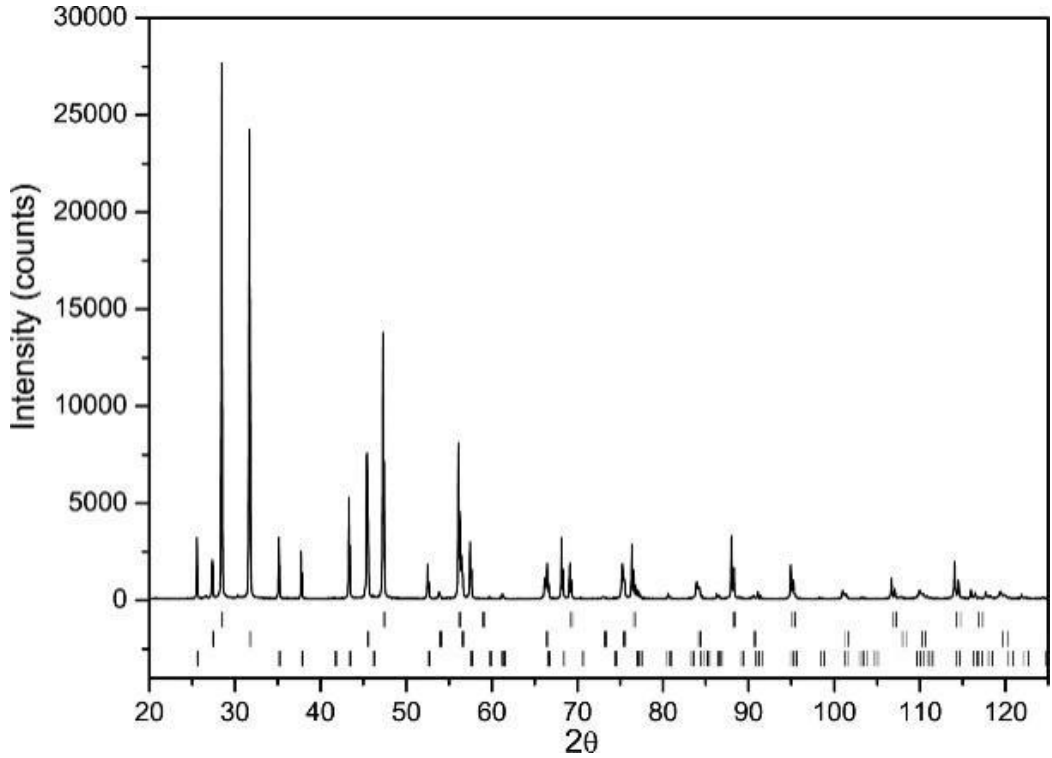

**Figure 1** X-ray powder diffraction pattern of a mixture of Si, NaCl and α-Al$_2$O$_3$ with the weight ratio of 1:1:1. Vertical short bars from top to bottom line indicate the Bragg positions of Si, NaCl and α-Al$_2$O$_3$, respectively.

A mixture of Si, NaCl and α-Al$_2$O$_3$ with the weight ratio of 1:1:1 was prepared, and the X-ray powder diffraction data of the mixture was collected using a Bruker D8 Advance diffractometer, which is operated in Bragg-Brentano geometry and equipped with Cu Kα irradiation. The powder pattern was shown in Figure 1. Equation (17) was applied to analyze the powder diffraction data and deduce the weight contents of the component phases. The intensity of each reflection of each component phase, namely $I_{jh}$ in equation (17), was retrieved from the powder pattern using pattern decomposition techniques. Although the experimental data was collected in the 2θ range of 20-125 °, the upper limit of 2θ was set to 80, 100 and 120 °, respectively, in the quantitative phase analysis to illustrate the effect of the upper limit of 2θ on the quality of quantitative phase analysis. The atomic structure data of Si and NaCl was used to calculate $\sum_h |F_h|^2$, while α-Al$_2$O$_3$ was treated as a structurally unknown phase, and its chemical contents in the unit cell were used to calculate $\sum_{i=1}^{n} \sum_{h'} f_{i,h'}^2$. All possible reflections below the upper limit of 2θ the in the calculation, and atomic displacements were not taken into account. The results of the quantitative phase analysis were presented in Table 2. The phase contents deduced using Rietveld's whole profile fitting method was also listed in Table 2 for comparison. In this case study, the quality of quantitative analysis using both methods seems to be comparable, as indicated by the similar deviations of deduced weight fractions from the "true" value, 33.3%. Nevertheless, the weight fractions deduced using our method fluctuate with the upper limit of $2\theta$ more greatly than the values obtained with whole profile fitting techniques. This characteristic reflects the difference between these two methods in the fundamental: the method proposed in this study quantifies the target phase content in a mixture using the sum of integrated intensity in a certain range of $2\theta$, while whole profile fitting technique measures the quantity of a component phase using the scale factor of the target phase's profile. The scale factor of the target phase' profile, theoretically, will not change with the range of $2\theta$, but the sum of the integrated intensity of the target phase will change greatly with the range of $2\theta$. In principle, when sufficient reflections are included in the calculation, the phase contents deduced by our method will converge to the "true" value. The example given here indicates that an X-ray powder diffraction pattern with an upper limit of $2\theta = 80°$ (for Cu K$\alpha$ irradiation) seems to be adequate for quantitative phase analysis. In comparison with the whole profile fitting techniques, the method proposed here has the advantage that quantitative phase analysis can be performed when only the chemical contents in the unit cell of the component phase are known, and one does not have to know the atomic structure of all component phases.

**Table 1** The results of quantitative phase analysis on a mixture of Si, NaCl and α-Al$_2$O$_3$ with the weight ratio of 1:1:1.

| Method | The upper limit of 2θ (°) | Weight fraction (%) | | |
|---|---|---|---|---|
| | | Si | NaCl | α-Al$_2$O$_3$ |
| Rietveld | 80 | 34.6 | 36.2 | 29.2 |
| | 100 | 34.7 | 35.7 | 29.6 |
| | 120 | 34.0 | 35.8 | 30.2 |
| This work | 80 | 36.2 | 31.7 | 32.1 |
| | 100 | 36.4 | 31.5 | 32.1 |
| | 120 | 35.3 | 29.1 | 35.6 |

**Acknowledgements**   The authors would like to thank Ms. WANG Qingqing of the Tianmu Lake Institute of Advanced Energy Storage Technologies for collecting the X-ray powder diffraction data.